\documentclass[twoside]{sfm_modified}
\usepackage{ulem}

\input{sf_modified.def}

\begin{document}

\def\llm{{\sc LLmodels}}
\def\atl{{\sc ATLAS9}}
\def\aatl{{\sc ATLAS12}}
\def\starsp{{\sc STARSP}}
\def\aur{$\Theta$~Aur}
\def\logg{\log g}
\def\tauros{\tau_{\rm Ross}}
\def\kms{km\,s$^{-1}$}
\def\bz{$\langle B_{\rm z} \rangle$}
\def\degr{^\circ}
\def\aaps{A\&AS}
\def\aap{A\&A}
\def\apjs{ApJS}
\def\apj{ApJ}
\def\rmxaa{Rev. Mexicana Astron. Astrofis.}
\def\mnras{MNRAS}
\def\actaa{Acta Astron.}
\newcommand{\Tef}{T$_{\rm eff}$~}
\newcommand{\Vt}{$V_t$}
\newcommand{\CC}{$^{12}$C/$^{13}$C~}
\newcommand{\CDC}{$^{12}$C/$^{13}$C~}

\thispagestyle{titlehead}

\setcounter{section}{0}
\setcounter{figure}{0}
\setcounter{table}{0}

\markboth{Mowlavi et al.}
         {New late B- and early A-type periodic variable stars in NGC~3766}

\titl{On the interpretation of new late B- and early A-type periodic variable stars in NGC~3766}
     {Mowlavi N., Saesen S., Barblan F., Eyer L.}
     {Geneva Observatory, University of Geneva, Switzerland\\
      Nami.Mowlavi@unige.ch}

\abstre{
We investigate possible interpretations of the new periodic B- and A-type variable stars discovered in NGC~3766.
They lie in the region of the Hertzsprung-Russell diagram between slowly pulsating B and $\delta$~Sct stars, a region where no pulsation is predicted by standard models of pulsating stars.
We show that the two other possible causes of periodic light curve variations, rotational modulation and binarity, cannot provide a satisfactory explanation for all the properties observed in those stars either.
The question of their origin is thus currently an open issue.
}

\baselineskip 12pt

\section{Introduction}
\label{Sect:introduction}

Asteroseismology has proven to be a unique tool to probe the interior of stars.
Certain conditions must be met for a star to pulsate, which translate into the existence of instability strips in the Hertzsprung-Russell (HR) diagram (Pamyatnykh 1999, Christensen-Dalsgaard 2004).
When a star is observed to pulsate, asteroseismology may provide information on the physical conditions inside the stars by comparing the observed frequencies with model predictions of pulsating stars.
An example of a breakthrough in main-sequence stars, besides the well known case of the Sun, is given by $\beta$~Cep stars, for which information on their interior rotation and convective core could be derived for a few of them (see, e.g., Sect.~7.3 in Aerts et al. 2010).


The classical instability strip of the $\kappa$~mechanism acting on H and He affects A- and early F-type main-sequence stars.
The corresponding pulsating stars are the $\delta$~Sct stars, which are characterized by a rich frequency spectrum containing up to several hundred frequencies from space-based observations.
The next instability strips on the main sequence are predicted at much higher luminosities in B-type stars, for which the $\kappa$~mechanism acts on the iron-group elements for slowly pulsating B (SPB) stars and $\beta$~Cep stars.
In between the $\delta$~Sct and SPB stars, an `instability gap' exists, where no pulsation is predicted to be sustained in the star.

Stars in this `gap' in the HR diagram are thus not expected to show periodic variability resulting from pulsations.
It is in this context that the discovery of periodic photometric variability in 36 such stars in the open cluster NGC~3766 has come as a surprise.
The origin of the periodic variability is currently unknown, but a clue is available from published spectra for four of them.
We summarize in Sect.~\ref{Sect:properties} their basic characteristics, and discuss possible origins of their variability in Sect.~\ref{Sect:origin}.
Conclusions are drawn in Sect.~\ref{Sect:conclusions}.

\section{Properties of the new early A- and late B-type variables}
\label{Sect:properties}


The properties of the late B- and early A-type periodic variables discovered in NGC~3766 are summarized in Sect.~7.1 of Mowlavi et al. (2013).
Those stars are distinguished from both the brighter SPB and the fainter $\delta$~Sct stars, not only by their location in the HR diagram --lying in the instability `gap' between those two classes of periodic variables--, but also by their frequencies.
The majority of the new variables have periods between 0.1 and 0.7 days, while SPB stars have periods larger than 0.5~days and $\delta$~Sct stars periods shorter than 0.25~days.

The variability amplitudes of the new variable stars are of the order of few milli-magnitudes or smaller.
About one third of them are found to be multi-periodic, to our variability detection limit of 1 to 2~mmag.
They are moreover stable on a time scale of at least seven years, the duration of the photometric monitoring campaign that led to their discovery.

Twenty percent of all stars in the instability `gap' are found to be periodic variables in NGC~3766.
The distribution of this percentage along the main sequence depends on the luminosity, starting with 40\% at the bright end, close to the region of SPB stars, and falling to 10\% at the faint end, close to $\delta$~Sct stars (see Fig.~2 in Mowlavi et al. 2014).
From the distribution of the variability amplitude, more stars in the considered magnitude range are actually expected to be variable, with amplitudes below the milli-magnitude level (see Mowlavi et al. 2013).

Finally, it is important to mention that four stars in our sample of new variables, for which spectra are available in the literature, are fast rotators, with rotational velocities above half their critical velocities.
This is a clue that fast rotation may play a key role in explaining the origin of the periodic variabilities.

\section{Possible origins of the periodic variability}
\label{Sect:origin}

Three scenarios can be invoked to explain periodic light variability in late B- and early A-type variables:
stellar pulsation, rotational modulation, and non-spherical stellar surface geometry due to the presence of a companion star in a binary system.
Each of these scenarios is considered in turn in the next subsections in light of the properties of the new periodic variables.

\subsection{Pulsation}
\label{Sect:pulsation}

A pulsation origin of the variability is suggested by the multi-periodicity of about one third of the new periodic variables, by the stability of the periodicity over at least seven years, and by the decrease of the fraction of periodic variables from late B- to early A--type stars (see below).

The periods and amplitudes of the new variables in NGC~3766 are very similar to those observed in CoRoT B-type targets reported in Degroote et al. (2009).
Those authors also claimed the existence of a new class of variable stars in the instability `gap' on the main sequence, based on the frequency spectra displayed by their B-type variables.
It must however be said that spectral type identification of CoRoT stars is not straightforward, and uncertainty exists in their stellar characterization that prevents a firm conclusion.

The main problem of the pulsation scenario is the fact that the periodic variables in NGC~3766 lie outside the instability strips predicted by standard models of pulsating stars.
This would act against a pulsational origin of the observed variability.

Non-standard models of pulsating stars can lead to other conclusions, though.
Rotation, for example, is known to alter the pulsation properties (e.g. Townsend 2005), but no reliable prediction for stars rotating at more than half their critical velocity exists yet.
The observation of fast rotation in all four of our new variables for which spectra are available in the literature supports this conclusion.
Four stars among 36 constitute small number statistics, but it suggests that fast rotation may play an important role in explaining those periodic variables.
It must also be said that the cluster is known to harbor many Be stars, which are commonly associated with fast rotation.

\subsection{Rotational modulation}
\label{Sect:spots}

Spots at the surface of rotating stars can also lead to photometric modulation.
Moreover, differential rotation could explain multi-periodic signals, to a certain extent, for suitable frequency separations.

Rotational modulation caused by star spots or some other co-rotating structure is favored by Balona (2013) to explain periodic light variations observed in Kepler A-type stars.
His conclusion is based on the similarity between the distribution of equatorial velocities of Kepler A-type stars (assuming equality between their photometric and rotation periods) and the distribution of equatorial velocities of A-type main-sequence field stars.
The fact that 1.5\% of Kepler A-type stars show flares in their light, indicating the presence of stellar activity in those stars, seems to further support his conclusion.

It must however be said that spectral type identification of Kepler stars is not easier than that of CoRoT stars.
Besides, their periods have a different distribution than those in NGC~3766 (compare his Fig.~4 with Fig.~19 in Mowlavi et al.~2013).
An origin of our new periodic variables attributed to spotted stars seems thus less likely, though it should not be excluded at this stage.
The amplitudes of variability reported by Balona (2013) are smaller than 0.2~mmag for 80\% of Kepler A-type periodic variables, which is below our photometric detection limit.
However, Balona (2013) reports on the presence of a tail in the amplitude distribution of Kepler A-type stars, with a fraction of stars having amplitudes larger than 1~mmag, the origin of which is unclear.

Another interesting study related to spotted stars is provided by Degroote et al. (2011), who focuss on several late B-type CoRoT targets (B8-B9 types) that show periodic variability.
The authors analyze and discuss to some length both pulsation and rotational modulation mechanisms.
They conclude that the observed frequencies are compatible with a spotted star in one case (for which they detect amplitudes of variability of the dominant modes between 1 and 2~mmag), while gravity mode pulsations are favored in four other cases (for which the amplitudes are less than 0.3~mmag). 

The main challenge of the spotted stars scenario is that no activity is expected in late B- and early A-type stars, due to the absence of an extended convective region below their surface.
Periodic variability is observed in a sub-class of chemically peculiar Ap and Bp stars having strong (fossil) magnetic fields (e.g., Mathys et al. 1989; see also other proceedings in the conference).
In those stars, diffusion of elements below their surface is modulated by the magnetic field, leading to the presence of spots at their surface that can explain the photometric variability.
However, they cannot be fast rotators for diffusion to operate.

The difficulties of the spotted star scenario are also recognized by Degroote et al. (2011) and Balona (2013) in their interpretation of the variability observed in late B- and A-type CoRoT and Kepler stars, respectively.
But we cannot strictly speaking exclude this interpretation for at least a fraction of the new variables in NGC~3766.

\subsection{Binarity}
\label{Sect:ellipsoidal}

Close binary systems (not necessarily eclipsing) can display periodic photometric variability if the surface of one or both components is deformed by tidal effects due to the gravitational influence of the close companion.
These stars are denominated as ellipsoidal binaries, and their light curves resemble a sinusoidal light curve.

This scenario must be considered seriously knowing that the fraction of binary systems is known to be high among massive stars.
But it can, at best, only explain the mono-periodic variables and not the multi-periodic ones.
If all mono-periodic variables turn out to be ellipsoidals, however, it would raise another question, i.e. why ellipsoidals with such periods are limited to the magnitude range between those of $\delta$~Sct and SPB stars.

\section{Conclusions}
\label{Sect:conclusions}

The strength of our observations resides in the fact that the stars belong to a cluster.
Therefore, their spectral types and their luminosities relative to one another can be easily known from their positions in the color-magnitude and color-color diagrams.
This secures the spectral types of the new periodic variables discovered in NGC~3766.

Our sample of new periodic late B- and early A-type stars may contain different types of variable stars.
Both pulsation and rotational modulation scenarios, the only viable ones for the multi-periodic candidates among them, present challenges for astrophysicists.
Rotational modulation requires the formation of spots at the surface of the stars, but no activity is expected for late B- and early A-type stars.
Pulsation, on the other hand, is not predicted for those types of stars by standard models, but non-standard models, such as models with very fast rotation, may change this conclusion.
In that sense, we cannot say that our observations are not compatible with predictions, since there is currently no such prediction.
Future work should concentrate, observationally on better characterizing the properties of the new variables, and theoretically on providing pulsation predictions in very fast rotating stars.



\begin{thebibliography}{}

Aerts, C., Christensen-Dalsgaard, J., \& Kurtz, D. W. 2010, \textit{Asteroseismology} (Heidelberg: Springer)\\

Balona, L. A. 2013, \textit{MNRAS} 431, 2240\\

Christensen-Dalsgaard, J. 2004, in:  \textit{SOHO 14 Helio- and Asteroseismology: Towards a Golden Future}, D. Danesy (ed.), ESA SP 559, p.\,1\\

Degroote, P., Acke, B., Samadi, R., et al. 2011, \textit{A\&A}, 536, A82\\

Degroote, P., Aerts, C., Ollivier, M., et al. 2009, \textit{A\&A} 506, 471\\

Mathys, G., Maitzen, H. M., North, P. et al. 1989, \textit{The Messenger} 55, 41\\

Mowlavi, N., Barblan, F., Saesen, S. \& Eyer, L. 2013, \textit{A\&A} 554, A108\\

Mowlavi, N., Saesen, S., Barblan, F. \& Eyer, L. 2014, in: \textit{Precision Asteroseismology}, W. Chaplin et al. (Eds), IAU Symp.~301, in press (arXiv: 1309.6612)\\

Pamyatnykh, A.\,A. 1999, \textit{Acta Astron.} 49, 119\\

Townsend, R. H. D. 2005, \textit{MNRAS}, 364, 573

%
%
%
%
%
%
%

\end{thebibliography}
\end{document}